\documentclass[conference,10pt]{IEEEtran}
\usepackage{epsfig,rotating,setspace,latexsym,amsmath,epsf,amssymb,amsfonts,bm,theorem,subfigure,epstopdf}
\usepackage{cite,authblk}
\usepackage{bbm}
\usepackage[ruled,vlined]{algorithm2e}
\DeclareMathOperator*{\argminA}{arg\,min}
\DeclareMathOperator*{\argmaxA}{arg\,max}
\usepackage{color}

\SetArgSty{textnormal}

\IEEEoverridecommandlockouts
\allowdisplaybreaks

\begin{document}
\bstctlcite{IEEEexample:BSTcontrol} 



\title{Covert Communications via Adversarial Machine Learning and Reconfigurable Intelligent Surfaces}
	
\author[1]{Brian Kim}
\author[2]{Tugba Erpek}
\author[2]{Yalin E. Sagduyu}
\author[1]{Sennur Ulukus}

\affil[1]{\normalsize Department of Electrical and Computer Engineering, University of Maryland, College Park, MD 20742, USA}
\affil[2]{\normalsize Intelligent Automation, Inc., Rockville, MD 20855, USA  \thanks{This effort is supported by the U.S. Army Research Office under contract W911NF-20-C-0055. The content of the information does not necessarily reflect the position or the policy of the U.S. Government, and no official endorsement should be inferred.}}

\maketitle
\begin{abstract}
By moving from massive antennas to antenna surfaces for software-defined wireless systems, the reconfigurable intelligent surfaces (RISs) rely on arrays of unit cells to control the scattering and reflection profiles of signals, mitigating the propagation loss and multipath attenuation, and thereby improving the coverage and spectral efficiency. In this paper, covert communication is considered in the presence of the RIS. While there is an ongoing transmission boosted by the RIS, both the intended receiver and an eavesdropper individually try to detect this transmission using their own deep neural network (DNN) classifiers. The RIS interaction vector is designed by balancing two (potentially conflicting) objectives of focusing the transmitted signal to the receiver and keeping the transmitted signal away from the eavesdropper. To boost covert communications, adversarial perturbations are added to signals at the transmitter to fool the eavesdropper's classifier while keeping the effect on the receiver low. Results from different network topologies show that adversarial perturbation and RIS interaction vector can be jointly designed to effectively increase the signal detection accuracy at the receiver while reducing the detection accuracy at the eavesdropper to enable covert communications.     
\end{abstract}

\section{Introduction}\label{sec:Introduction}
Reconfigurable intelligent surfaces (RISs) have emerged as novel tools for software-defined wireless communications to increase the coverage and the spectral efficiency of 5G and beyond wireless communication systems \cite{Akyildiz1, basar, hu2018beyond}. RISs correspond to large number of reflecting antennas that can be controlled to interact with incident signals. Specifically, the phase shifts of the RISs can be controlled without the need of any computing or energy source for decoding, encoding, or transmission. For that purpose, it is necessary to select the best reflection beamforming or interaction vector at the RIS to focus the incident beam towards the receiver. However, this is a complex task as reflection properties (as in phase shifts) need to be optimized for a large number of antenna elements.

By effectively learning from rich representations of spectrum data, deep learning (DL) has found a broad set of applications including signal classification, waveform design, and wireless security \cite{erpek1}. DL has been effectively applied to solve the complex task of optimizing the RIS-aided communications. The interaction vector at the RIS was designed in \cite{taha} by using the channel information as the input to the deep neural network (DNN). Reinforcement learning (RL) was applied in \cite{taha3} to predict the interaction vector at the RIS without the need for an external source to determine it. A recurrent neural network was used in \cite{nof} to predict whether to use a direct link or the RIS, and in the latter case to predict the best RIS beam. For indoor communications, DL was used in \cite{huang2019indoor} for the RISs to improve the focus of transmitted signals to receiver positions. Joint design of transmit beamforming at the base station and phase shift at the RIS was studied in \cite{huang2020reconfigurable} to maximize the sum rate of multiuser downlink MIMO systems with deep RL. A convolutional neural network (CNN) was used in \cite{yang2021intelligent} to identify the interfering users from the incident signal at the RIS. The RIS was integrated with autoencoder communications in \cite{RIS_AE} by training the DNNs for the RIS, the encoder at the transmitter, and the decoder at the receiver. 

While DL has been instrumental to achieve the promising benefits of the RIS, DL itself is susceptible to attacks. In general, the DNNs are known to be highly vulnerable to adversarial perturbations added to the inputs of the DNNs to induce an incorrect output or misclassification result \cite{goodfellow2014explaining}. Adversarial attacks have been first introduced in the computer vision domain \cite{Szegedy1} and then later extended to other domains as DL finds new applications. There are various forms of attacks launched against the DNNs, collectively studied under adversarial machine learning (AML). Due to the shared and open nature of the wireless medium, AML attacks can be launched over the air to target the victim DNNs used for wireless communication applications. Recently, AML attacks have been studied as the emerging threat to wireless security \cite{Sagduyu2020, AML_RF}. Different types of attacks have been considered such as exploratory (inference) attacks \cite{Terpek}, adversarial (evasion) attacks \cite{Larsson2, Kokalj2, Kokalj3, Flowers1, Bair1, Lin, Kim1, Kim2, KimMultiple, KimICC, Kimpower}, poisoning (causative) attacks \cite{Sagduyu1}, membership inference attacks \cite{MIA}, and Trojan attacks \cite{Davaslioglu1}. AML can also support covert communications by fooling the DNN-based signal classifiers of eavesdroppers \cite{Gunduz1, Gunduz2, Kim5G}.

In this paper, we consider RIS-aided wireless communications, where a receiver uses its DNN classifier to detect the transmitter's signal that is reflected by the RIS. Concurrently, there exists an eavesdropper with another DNN classifier to identify an ongoing transmission for adversarial purposes. The transmitter adds adversarial perturbations to its signals to fool the eavesdropper and reduce its detection accuracy. Minimum power is used for these adversarial perturbations to minimize the effect on the receiver's detection performance. Simultaneously, the RIS interaction vector (corresponding to phase shifts induced by the RIS antenna elements) is designed so that the RIS reflects the signal towards the intended receiver while keeping the reflection away from the eavesdropper. 

Note that the prior work on the RISs has typically considered improving the performance (such as the signal-to-noise-ratio (SNR) at the receiver) by optimizing the RIS interaction vector only for the receiver. First, we show that this approach does not guarantee covertness of the signals at the eavesdropper. In particular, while the SNR of the receiver is highly correlated with the receiver's detection accuracy and can effectively guide the optimization of the RIS interaction vector to maximize the receiver performance, it does not reliably tell how to design the RIS to reduce the eavesdropper's detection accuracy. Then, we show how to reduce the eavesdropper's performance by adding adversarial perturbations to the transmitter signals that are reflected by the RIS. For that purpose, we consider different topologies and analyze how the design of the RIS interaction vector for covert communications adapts to different locations of the receiver and the eavesdropper. Our results show that the beam selection of the RIS is the crucial component for covert communications when the transmitter has low power budget for adversarial attack. However, when there is enough power budget, the adversarial perturbation becomes the dominant factor to improve covert communications.

The rest of the paper is organized as follows. Section~\ref{sec:SystemModel} presents the system model. Section~\ref{sec:Adversarial} describes adversarial perturbations for covert communications. Section~\ref{sec:Performance} specifies the topology and channel models, and presents the performance results. Section~\ref{sec:Conclusion} concludes the paper. 

\section{System Model} \label{sec:SystemModel}
We consider a communication system where a transmitter is transmitting the signal $x$ while the intended receiver uses a pretrained DNN classifier to detect the ongoing signal that is reflected by the RIS equipped with $N$ reconfigurable antenna elements. The transmitter and the intended receiver have a single antenna each. Concurrently, there exists an eavesdropper with a single antenna that also aims to detect the ongoing signal using another pretrained DNN classifier. 
To defend against eavesdropping, the transmitter adds a perturbation $\delta$ to its signal, which corresponds to an adversarial attack to the eavesdropper.
We describe in Section~\ref{sec:Adversarial} how to craft this perturbation for covert communications. In addition, we design the RIS interaction vector $\boldsymbol{\psi}$ so that the adversarial attack becomes most effective on the eavesdropper while minimizing the effect on the classifier at the intended receiver. In other words, the designs of the adversarial perturbation and the RIS interaction vector are coupled, and should be jointly performed. We assume that the RIS interaction vector $\boldsymbol{\psi}$ is selected from a predefined codebook $\mathcal{S}$.


When the transmitter transmits $x$, the input to the RIS (namely, the incident signal for the RIS) is given by
\begin{equation}\label{eq:ris}
	\boldsymbol{x}_{ris}(x) = \boldsymbol{h}_{tr}x, 
\end{equation}
where $\boldsymbol{h}_{tr}\in \mathbb{C}^{N \times 1}$ is the channel between the transmitter and the RIS. We assume that the phase shift of the RIS element is quantized and represented with 1 bit where each RIS element introduces either $0^{\circ}$ or $180^{\circ}$ phase shift and $\kappa$ loss to the signal. Thus, the signal at the output of RIS is given by
\begin{equation}\label{eq:output}
	[\boldsymbol{y}_{ris}(x)]_{i} = c_{i}[\kappa\boldsymbol{x}_{ris}(x)]_{i},  \quad i = 1,\cdots, N, 
\end{equation}
where $c_{i}\in \{-1,1\}$ or $c_{i}=e^{j\theta_i}$ and $\theta_i$ corresponds to the phase shifts (e.g., $\theta_{i} \in \{0,\pi\}$). No noise is added at the RIS (in accordance with previous RIS studies) since it is a passive device. The received signal at the intended receiver is 
\begin{equation}\label{eq:receive}
	y_{r}(x) = \boldsymbol{h}_{ri}^{T}\boldsymbol{y}_{ris}(x)+n_r,
\end{equation}
where $n_r$ is the noise at the intended receiver and $\boldsymbol{h}_{ri}\in \mathbb{C}^{N \times 1}$ is the channel between the RIS and the intended receiver. This channel formulation takes the channel gain and the phase shift between the RIS and the intended receiver into account. Simultaneously, the eavesdropper receives the signal
\begin{equation}\label{eq:eve}
	y_{eve}(x) = \boldsymbol{h}_{re}^{T}\boldsymbol{y}_{ris}(x)+n_{e},
\end{equation}
where $n_{e}$ is the noise at the eavesdropper and $\boldsymbol{h}_{re}\in \mathbb{C}^{N \times 1}$ is the channel between the RIS and the eavesdropper (taking the channel gain and the phase shift between the RIS and the eavesdropper into account). When the transmitter transmits $x+\delta$, the input to the RIS expression of (\ref{eq:ris}) changes to 
\begin{equation}\label{eq:ris2}
	\boldsymbol{x}_{ris}(x+\delta) = \boldsymbol{h}_{tr}x+\boldsymbol{h}_{tr}\delta,
\end{equation}
and (\ref{eq:output}), (\ref{eq:receive}), and (\ref{eq:eve}) change accordingly.

\begin{figure}[t]
	\centerline{\includegraphics[width=0.925\linewidth]{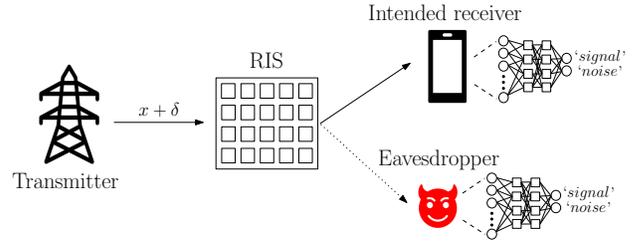}}
	\caption{RIS-aided communications in the presence of an eavesdropper.}
	\label{plot:ris}
\end{figure}

We define the pretrained classifier at the intended receiver as $f_{r}(.;\boldsymbol{\theta}_{r}): \mathcal{X} \rightarrow \mathbb{R}^{2}$, to determine the existence of ongoing background transmission to utilize the idle bands, where  $\boldsymbol{\theta}_{r}$ is the set of transmitter's DNN parameters and $\mathcal{X} \subset \mathbb{C}^{M}$. Note that the input to the DNN is defined as $\boldsymbol{y}_{r}(\boldsymbol{x}) = [y_{r}(x_1), y_{r}(x_2), \cdots, y_{r}(x_M)]\in \mathcal{X}$. The input $\boldsymbol{y}_{r}(\boldsymbol{x})$ is assigned to the label $\hat{l}_{r}(\boldsymbol{y}_{r}(\boldsymbol{x}),\boldsymbol{\theta}_{r}) = \arg \max_{q} f^{(q)}_{r}(\boldsymbol{y}_{r}(\boldsymbol{x}),\boldsymbol{\theta}_{r})$, where $f_{r}^{(q)}(\boldsymbol{y}_{r}(\boldsymbol{x}),\boldsymbol{\theta}_{r})$ is the output of classifier $f_{r}^{(q)}$ corresponding to the $q$th class. Concurrently, the eavesdropper tries to detect the background transmission based on the $\boldsymbol{y}_{eve}(\boldsymbol{x})= [y_{eve}(x_1), y_{eve}(x_2), \cdots, y_{eve}(x_M)]\in \mathcal{X}$ using its own DNN classifier. We define the classifier of the eavesdropper as $f_{eve}(.;\boldsymbol{\theta}_{eve}): \mathcal{X} \rightarrow \mathbb{R}^{2}$. The input $\boldsymbol{y}_{eve}(\boldsymbol{x})$ is assigned to the label $\hat{l}_{eve}(\boldsymbol{y}_{eve}(\boldsymbol{x}),\boldsymbol{\theta}_{eve}) = \arg \max_{q} f^{(q)}_{eve}(\boldsymbol{y}_{eve}(\boldsymbol{x}),\boldsymbol{\theta}_{eve})$, where $f_{eve}^{(q)}(\boldsymbol{y}_{eve}(\boldsymbol{x}),\boldsymbol{\theta}_{eve})$ is the output of classifier $f_{eve}^{(q)}$ corresponding to the $q$th class.

\section{Adversarial Attack against Eavesdropping} \label{sec:Adversarial}
In this section, we introduce how to design an adversarial attack against the eavesdropper to cause misclassification. Since the adversarial perturbation that is transmitted at the transmitter is reflected by the RIS before it is received at the eavesdropper, we need to take the RIS interaction vector into account while generating the adversarial attack. If the transmitter transmits $x+\delta$, then the eavesdropper receives
\begin{equation}\label{perturbation2}
	y_{eve}(x+\delta) =  \boldsymbol{h}_{re}^{T}\boldsymbol{\Phi}\boldsymbol{h}_{tr}(x+\delta)+n_{e},
\end{equation}
where $\boldsymbol{\Phi} = \text{diag}[\phi_1, \phi_2, \cdots, \phi_N] \in \mathbb{C}^{N \times N}$ and $\phi_k = c_k\kappa$. 

The transmitter designs the adversarial perturbation $\boldsymbol{\delta}$ to cause misclassification at the eavesdropper while limiting its effect on the intended receiver by designing the RIS interaction vector simultaneously. Thus, the transmitter determines $\boldsymbol{\delta}$ by solving the following optimization problem:
\begin{align} \label{perturbation}
	\argminA_{\boldsymbol{\delta}}& \quad ||\boldsymbol{\delta}||_2\nonumber\\
	\mbox{s.t.} \quad&\hat{l}_{eve}(\boldsymbol{y}_{eve}(\boldsymbol{x}),\boldsymbol{\theta}_{eve}) \ne \hat{l}_{eve}(\boldsymbol{y}_{eve}(\boldsymbol{x}+\boldsymbol{\delta}),\boldsymbol{\theta}_{eve}) \nonumber\\
	 \quad & ||\boldsymbol{\delta}||^{2}_{2} \le P_{max}. 
\end{align}

However, (\ref{perturbation}) is hard to solve due to the nonconvexity of the DNN structure. Therefore, we use fast gradient method (FGM) \cite{goodfellow2014explaining} to linearize the loss function,  $L_{eve}(\boldsymbol{\theta}_{eve},\boldsymbol{y}_{eve}(\boldsymbol{x}),\boldsymbol{l})$, of the DNN in the neighborhood of input $\boldsymbol{y}_{eve}(\boldsymbol{x})$, where $\boldsymbol{l}$ is the label vector, and use the linearized loss function for the optimization. In this paper, we consider a targeted attack against the eavesdropper such that the transmitter designs the perturbation that decreases the loss function of the class `noise' to enforce a specific misclassification, from label `signal' to label `noise', at the eavesdropper by transmitting the perturbation in the opposite direction of the gradient of the loss function $-\nabla_{\boldsymbol{y}_{eve}(\boldsymbol{x})}L_{eve}(\boldsymbol{\theta}_{eve},\boldsymbol{y}_{eve}(\boldsymbol{x}),\boldsymbol{l}^{\textit{target}})$, where $\boldsymbol{l}^{\textit{target}}$ is `noise' class. However, the channels between the nodes change the direction of the attack $\boldsymbol{\delta}$ that is first sent at the transmitter.
Thus, the transmitter takes the effect of the channels and the RIS interaction vector into account by multiplying $(\boldsymbol{h}_{re}^{T}\boldsymbol{\Phi}\boldsymbol{h}_{tr})^*$ with the gradient of the loss function as it has been done similarly in \cite{Kim1}.
During the adversarial attack generation process, we assume that the transmitter has the information about all channels and the RIS interaction vector. Knowing the channel between the RIS and the eavesdropper is difficult for real systems. This assumption can be relaxed as in \cite{Kim2} to know the channel distribution between the RIS and the eavesdropper instead of the channel instance. The detailed algorithm is presented in Algorithm \ref{alg1}.

\begin{algorithm}[t]
	\DontPrintSemicolon
	\SetAlgoLined
	\label{alg1}
	Inputs: $\boldsymbol{y}_{eve}(\boldsymbol{x})$, desired accuracy $\varepsilon_{acc}$, power budget $P_{\textit{max}}$ and eavesdropper's DNN architecture \\
	Initialize: $\varepsilon_{\textit{max}} \leftarrow \sqrt{P_{\textit{max}}}, \varepsilon_{min} \leftarrow 0, \boldsymbol{l}^{\textit{target}} \leftarrow \text{`noise'}$ \\
	$\boldsymbol{\delta}_{norm} =\frac{(\boldsymbol{h}_{re}^{T}\boldsymbol{\Phi}\boldsymbol{h}_{tr})^*\nabla_{\boldsymbol{y}_{eve}(\boldsymbol{x})}L_{eve}(\boldsymbol{\theta}_{eve},\boldsymbol{y}_{eve}(\boldsymbol{x}),\boldsymbol{l}^{\textit{target}})}{(||(\boldsymbol{h}_{re}^{T}\boldsymbol{\Phi}\boldsymbol{h}_{tr})^*\nabla_{\boldsymbol{y}_{eve}(\boldsymbol{x})}L_{eve}(\boldsymbol{\theta}_{eve},\boldsymbol{y}_{eve}(\boldsymbol{x}),\boldsymbol{l}^{\textit{target}})||_{2})}$\\
	\If{$\hat{l}_{eve}(\boldsymbol{y}_{eve}(\boldsymbol{x}),\boldsymbol{\theta}_{eve})== \text{`signal'}$}{
		\While{$\varepsilon_{\textit{max}}-\varepsilon_{min} > \varepsilon_{acc}$}{
			$\varepsilon_{avg} \leftarrow (\varepsilon_{\textit{max}}+\varepsilon_{min})/2$\\
			$\boldsymbol{x}_{adv} \leftarrow \boldsymbol{y}_{eve}(\boldsymbol{x}) - \varepsilon_{avg}\boldsymbol{\delta}_{\textit{norm}}$\\
			\lIf{$\hat{l}_{eve}(\boldsymbol{x}_{adv},\boldsymbol{\theta}_{eve})==  \text{`noise'}$}{
				$\varepsilon_{min}\leftarrow \varepsilon_{avg}$}
			\lElse{$\varepsilon_{\textit{max}}\leftarrow \varepsilon_{avg}$}
	}}{$\varepsilon = \varepsilon_{\textit{max}}$, $\boldsymbol{\delta}^{*} = -\varepsilon\boldsymbol{\delta}_{norm} $}\\
	\caption{Crafting the adversarial attack at the transmitter against the eavesdropper.}
\end{algorithm}

The RIS interaction vector is determined from the predefined codebook $\mathcal{S}$ that maximizes the classifier accuracy at the intended receiver while minimizing the classifier accuracy at the eavesdropper. Denote the accuracy of the intended receiver's classifier as $P_{acc,i}(\boldsymbol{x})$ and the accuracy of the eavesdropper's classifier as $P_{acc,e}(\boldsymbol{x})$ when the transmitted signal is $\boldsymbol{x}$. Then, the RIS interaction vector is selected as
\begin{equation}\label{ris selection}
	\boldsymbol{\psi}^* = \argmaxA_{\boldsymbol{\psi}\in \mathcal{S}} P_{acc,i}(\boldsymbol{x}+\boldsymbol{\delta})-P_{acc,e}(\boldsymbol{x}+\boldsymbol{\delta}).
\end{equation}

\begin{figure}
    \centering
    \includegraphics[width=0.5\textwidth]{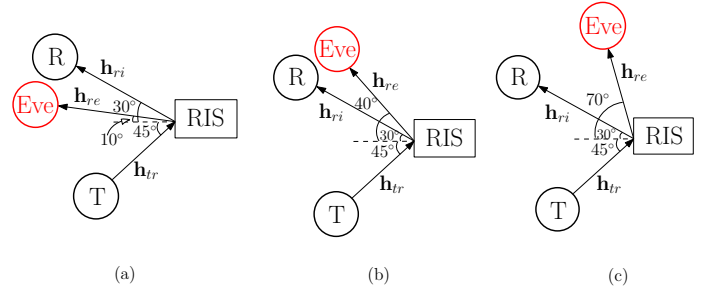}
    \caption{Different locations of the eavesdropper `Eve' while the locations of transmitter `T', receiver `R', and the RIS are fixed.}
    \label{fig:topology}
\end{figure}

		
	
		
		
	
	

\section{Performance Evaluation}\label{sec:Performance}
\subsection{Topology and Channel Models} \label{sec:Topology}
We consider different topologies to study the effect of the RIS on the classifier performance at the eavesdropper and assess how the location of the eavesdropper with respect to the location of the intended receiver affects the RIS interaction vector selection according to (\ref{ris selection}). We fix the location of the intended receiver while changing the location of the eavesdropper to analyze how the selection of the RIS interaction vector changes. We define the incident angle of the transmitted signal from the transmitter to the RIS as $\theta_{tr}$, the reflected angle from the RIS to the receiver as  $\theta_{ri}$, and the reflected angle from the RIS to the eavesdropper as  $\theta_{re}$. We set the angles $\theta_{tr} = 45^\circ$ and $\theta_{ri}=30^\circ$, and change the location of the eavesdropper by changing the reflected angle from the RIS to the eavesdropper as  $\theta_{re} = 10^\circ, 40^\circ, 70^\circ$, as shown in Fig. \ref{fig:topology}. 
We define channels $\boldsymbol{h}_{tr}, \boldsymbol{h}_{ri}, \text{and}\; \boldsymbol{h}_{re}$ according to the wideband geometric channel model adopted in \cite{taha}, where $\boldsymbol{h}_{tr}$ is given by
\begin{equation}\label{channel}
	\boldsymbol{h}_{tr} = \sqrt{{\rho_{tr}N}}\boldsymbol{a}(\theta_{tr}),	
\end{equation}
where the $\rho_{tr}$ is the path loss and $\boldsymbol{a}(\theta_{tr})$ is the array response vector of the RIS at the angles of arrival $\theta_{tr}$, which is defined as $\boldsymbol{a}(\theta_{tr}) = \sqrt{\frac{1}{N}}[1, e^{jd\cos(\theta_{tr})}, \cdots,  e^{jd(N-1)\cos(\theta_{tr})}]^{T}$. The channels $\boldsymbol{h}_{ri}$ and $\boldsymbol{h}_{re}$ are defined similarly. For performance evaluation, we set $N=16$ and the spacing between reconfigurable antenna elements to the half of the wavelength. The predefined codebook adopts a discrete Fourier transform (DFT) codebook used in \cite{taha}, where the $i$th codebook is defined as $\boldsymbol{\psi}_i = [1, e^{j2\pi i/N},\cdots, e^{j2\pi(N-1)i/N}]^T$.

\subsection{Deep Learning Classifiers}
We assume that the transmitter transmits QPSK signals to the RIS. The classifiers at the receiver and the eavesdropper are modeled as two (different) CNNs, where the input to each CNN is of two dimensions (2,16) corresponding to 16 in-phase/quadrature (I/Q) data samples. The classifier architecture used in the simulations consists of a convolutional layer with kernel size $(1,3)$, a hidden layer with dropout rate $0.1$, ReLu activation function at convolutional and hidden layers, and softmax activation function at the output layer that provides the label `signal' or `noise'. We use cross-entropy as the loss function of the CNN that is implemented in Keras with TensorFlow backend. To collect the dataset to train the classifier, we let the transmitter transmit the signals with power 30dBm that are reflected by all possible $K=16$ RIS interaction vectors, e.g., RIS 1, RIS 2, $\cdots$, RIS 16, and three different SNR levels, e.g., 3dB, 5dB and 7dB, to train the receiver at a specific location. We collect 5000 samples for each RIS interaction vector and SNR level, generating 240000 signal samples. In addition, we generate 240000 noise samples and obtain 480000 samples in total to train and validate the classifier. We use half of the samples for training and the other half for validating the classifier.

\begin{figure}[t]
	\centerline{\includegraphics[width=0.92\linewidth]{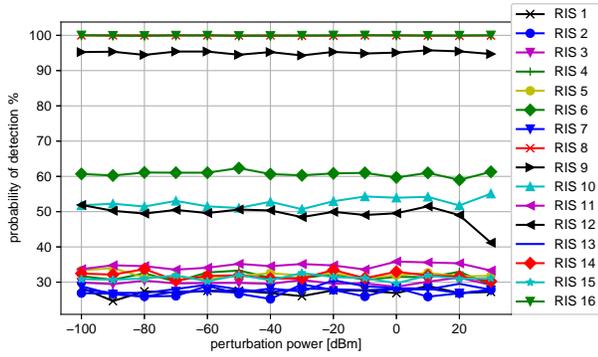}}
	\caption{Probability of detection for the classifier of the receiver when $\theta_{tr} = 45^\circ$, $\theta_{ri}=30^\circ$ and $\theta_{re} = 10^\circ$.}
	\label{rx100}
\end{figure}
\begin{figure}[t]
	\centerline{\includegraphics[width=0.92\linewidth]{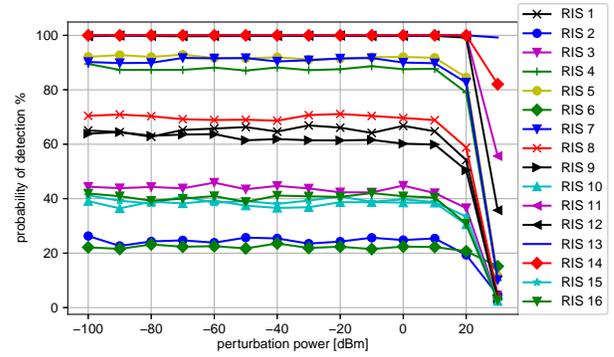}}
	\caption{Probability of detection for the classifier of the eavesdropper when $\theta_{tr} = 45^\circ$, $\theta_{ri}=30^\circ$ and $\theta_{re} = 10^\circ$.}
	\label{eve100}
\end{figure}

\subsection{Covert Communications Performance}

Once we train the classifiers for different locations of the eavesdropper and the receiver, we test the performance of the classifiers at the receiver and the eavesdropper when the transmitter transmits the signal with the adversarial perturbation added to fool the eavesdropper. During the test time, we fix the transmit power of the signal at the transmitter as 30dBm and set the noise power that results in an average of 5dB SNR at the receiver and the eavesdropper. Note that the power for adversarial perturbation at the transmitter is used separately from the transmit power of the signal.

We first motivate the need to design the RIS interaction vector differently when the eavesdropper is present. For that purpose, we investigate the correlation between the SNR and the probability of detection at the receiver, and the correlation between the SNR at the receiver and the probability of detection at the eavesdropper. We measure the correlation by the Pearson correlation coefficient. Without the eavesdropper, the RIS interaction vector is typically selected as the one that maximizes the SNR at the receiver. This selection is expected to yield a high probability of detection at the receiver. As an example, the correlation between the SNR and the probability of detection at the receiver is 0.94 for the topology shown in Fig. \ref{fig:topology}(c). This means that as expected, the SNR is a good measure to design the RIS for signal detection at the receiver. However, the correlation between the SNR at the receiver and the probability of detection at the eavesdropper is 0.69. This means that designing the RIS based on the SNR may be also good for the eavesdropper. Especially for high SNR at the receiver, the eavesdropper maintains moderate probability of detection. Therefore, we need a better criterion to select the RIS interaction vector than just selecting the one with highest SNR at the receiver, and additional means such as adversarial perturbation is needed to enable covert communications.

In this section, we investigate how different locations of the eavesdropper described in Section \ref{sec:Topology} affect the adversarial perturbation performance and the RIS interaction vector selection. 
First, we assess the performance of the classifier at the intended receiver with location given in Fig. \ref{fig:topology}(a). Fig.~\ref{rx100} shows that the classifier at the receiver is not affected by the adversarial perturbation even when its power is increased. Also, the probability of detection at the receiver differs significantly for different RIS interaction vectors. The probability of detection using RIS 16 and RIS 14 is 100\% and 30\%, respectively. From Fig. \ref{rx100}, the best RIS interaction vectors to select for the receiver are RIS 8 and RIS 16 leading to 100\% probability of detection at the receiver.

The performance of the classifier at the eavesdropper with location from Fig. \ref{fig:topology}(a) is presented in Fig. \ref{eve100}. The probability of detection at the eavesdropper also differs considerably with respect to different RIS interaction vectors. The adversarial perturbation reduces the probability of detection at the eavesdropper and causes misclassifications very likely when using more than 20dBm perturbation power at the transmitter. For different RIS interaction vectors, the adversarial perturbation has different degrees of effect on the eavesdropper's classifier. In particular, the adversarial perturbation through RIS 12 has more effect on the classifier than the adversarial perturbation through RIS 14. To determine the RIS interaction vector that induces the best probability of detection at the receiver while causing the worst performance at the eavesdropper, we select the RIS interaction vector based on (\ref{ris selection}) by analyzing Fig. \ref{rx100} and Fig. \ref{eve100}. For this topology, the best RIS interaction vector to use is RIS 16 and the probability of detection at the eavesdropper can drop to almost zero by jointly designing the RIS interaction vector and the adversarial perturbation.

\begin{figure}[t]
	\centerline{\includegraphics[width=0.92\linewidth]{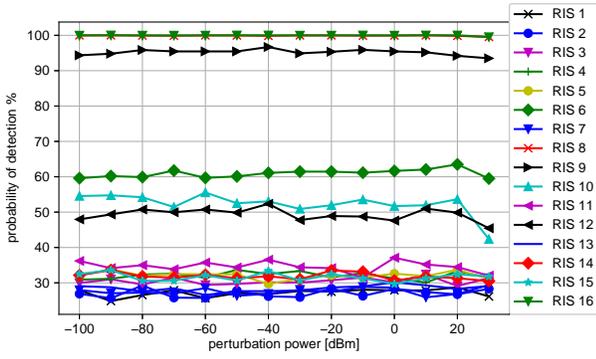}}
	\caption{Probability of detection for the classifier of the receiver when $\theta_{tr} = 45^\circ$, $\theta_{ri}=30^\circ$ and $\theta_{re} = 40^\circ$.}
	\label{rx130}
\end{figure}
\begin{figure}[t]
	\centerline{\includegraphics[width=0.92\linewidth]{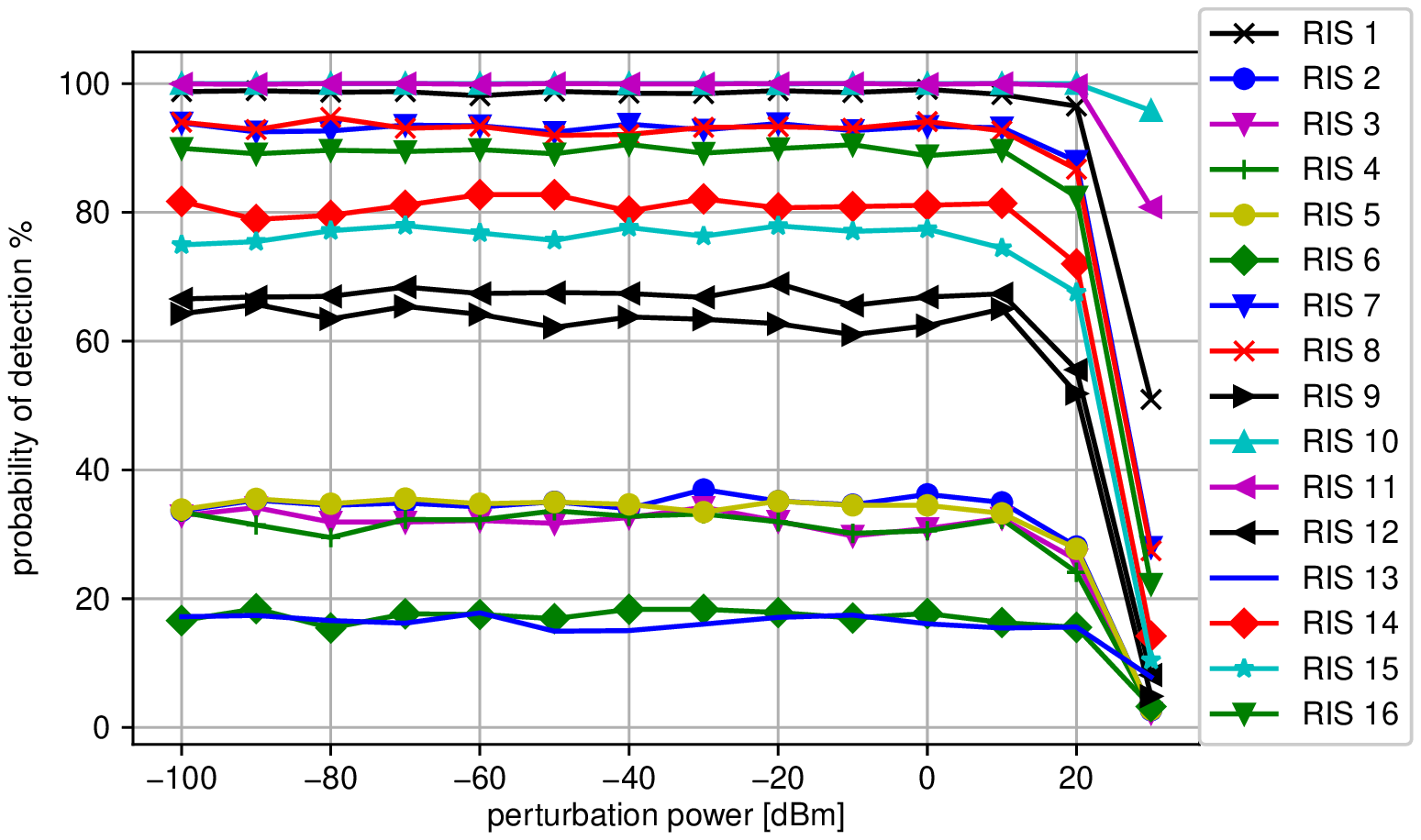}}
	\caption{Probability of detection for the classifier of the eavesdropper when $\theta_{tr} = 45^\circ$, $\theta_{ri}=30^\circ$ and $\theta_{re} = 40^\circ$.}
	\label{eve130}
\end{figure}
\begin{figure}[t]
	\centerline{\includegraphics[width=0.92\linewidth]{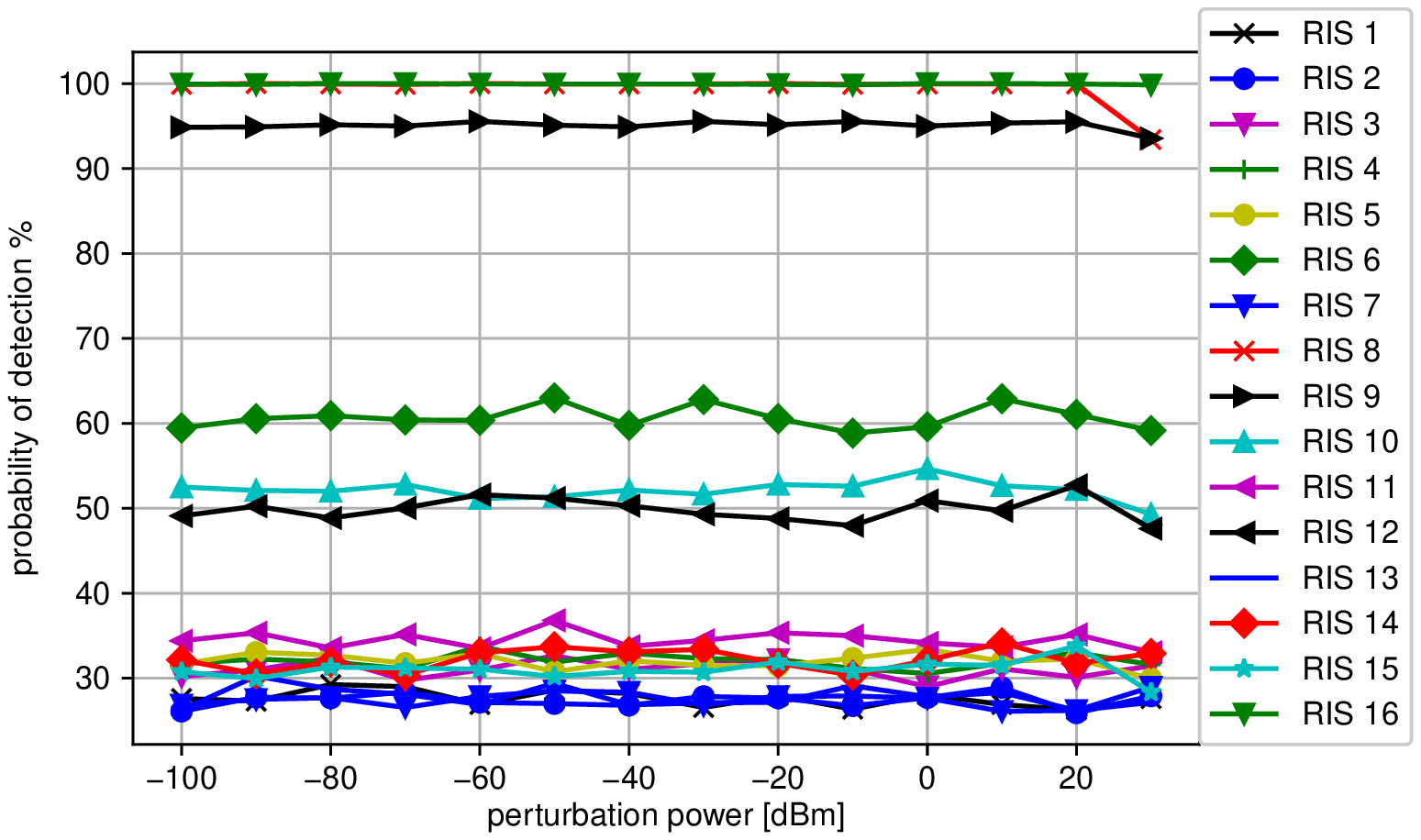}}
	\caption{Probability of detection for the classifier of the receiver when $\theta_{tr} = 45^\circ$, $\theta_{ri}=30^\circ$ and $\theta_{re} = 70^\circ$.}
	\label{rx160}
\end{figure}
\begin{figure}[t]
	\centerline{\includegraphics[width=0.92\linewidth]{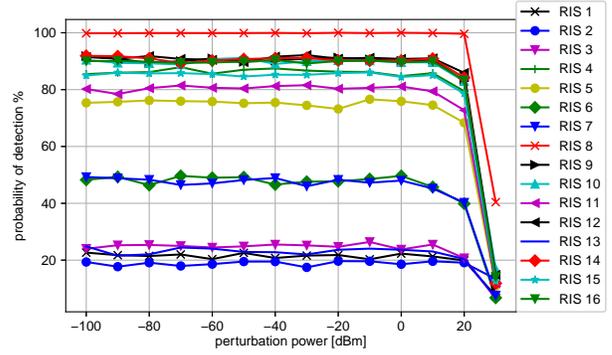}}
	\caption{Probability of detection for the classifier of the eavesdropper when $\theta_{tr} = 45^\circ$, $\theta_{ri}=30^\circ$ and $\theta_{re} = 70^\circ$.}
	\label{eve160}
\end{figure}

Next, we assess the performance for the topology given in Fig.~\ref{fig:topology}(b). The performance of the classifier at the receiver in Fig. \ref{rx130} is very similar to the result for the receiver in Fig. \ref{rx100}, since the location of the receiver is exactly the same. However, due to the change in location for the eavesdropper, it is observed in Fig. \ref{eve130} that the order of the RIS interaction vector from the highest probability of detection to the lowest has changed compared to Fig. \ref{eve100}. Again, to determine the best RIS interaction vector, we analyze Fig. \ref{rx130} and Fig. \ref{eve130}, and select the RIS interaction vector that provides the maximum value for the probability of detection difference between the receiver and the eavesdropper. For this topology, the best RIS interaction vectors for the receiver are RIS 8 and RIS 16 without considering the performance of the eavesdropper's classifier. However, RIS 8 and RIS 16 are also good for the eavesdropper, since the probability of detection at the eavesdropper for those RIS interaction vectors is around 90\%, yielding around 10\% difference between the probability of detection at the receiver and eavesdropper. Instead, for covert communications, we need to select RIS 9, which reduces the detection probability at the receiver to 95\% compared to 100\% for RIS 8 and RIS 16, but RIS 9 enforces the probability of detection at the eavesdropper to drop to 65\% without any adversarial perturbation. Furthermore, the probability of detection at the eavesdropper for RIS 9 can be reduced to 10\% by adding an adversarial perturbation at the transmitter.

Finally, we assess the performance for the topology given in Fig.~\ref{fig:topology}(c). The performance of the classifier at the receiver is similar to the performance in other topologies except that the probability of the detection for RIS 8 decreases when the adversarial perturbation is added with higher power. Fig. \ref{eve160} shows that the order of the RIS interaction vector from the highest probability of detection to the lowest has changed again compared to the order from other topologies. The best RIS interaction vectors at the receiver are RIS 8 and RIS 16 leading to 100\% probability of detection, but the probability of detection at the eavesdropper using RIS 16 and RIS 8 is 100\% and 90\%, respectively, without a perturbation. Thus, RIS 16 is chosen over RIS 8, but the eavesdropper still can detect the signal with 90\% accuracy. Adversarial perturbation is needed for covertness since the best RIS interaction vector for the receiver is also the best one for the eavesdropper. When RIS 16 is used, the probability of detection at the eavesdropper drops to 10\% when the transmitter uses 25dBm power for adversarial perturbation, while the probability of detection at the receiver remains 100\%.

\section{Conclusion}\label{sec:Conclusion}
We considered RIS-aided wireless communications where the receiver uses its DNN classifier to detect the ongoing transmission reflected by the RIS. Concurrently, there exists an eavesdropper that also tries to detect the ongoing transmission for its adversarial purposes. To make the communications covert, the transmitter crafts the adversarial perturbation to cause misclassifications at the eavesdropper. In addition, the RIS interaction vector that determines the direction of the reflected signal is designed so that the reflected signal is focused to the receiver while keeping it away from the eavesdropper. Through different topologies, we showed that the design of the RIS interaction vector for covert communications changes with respect to the location of not only the receiver and but also the eavesdropper. Moreover, the adversarial perturbation that is generated at the transmitter further improves the covertness of communications and has only a negligible effect on the receiver performance.

\section*{Acknowledgement}
We thank Prof. Ahmed Alkhateeb for the discussions on the RIS codebook generation.

\bibliographystyle{IEEEtran}
\bibliography{lib.bib}

\end{document}